\documentclass{article}

    \PassOptionsToPackage{numbers, compress}{natbib}

\usepackage[dblblindworkshop, final]{coml_2025}
\workshoptitle{Constrained Optimization for Machine Learning (COML)}

\usepackage[utf8]{inputenc} 
\usepackage[T1]{fontenc}    
\usepackage{hyperref}       
\usepackage{url}            
\usepackage{booktabs}       
\usepackage{amsfonts}       
\usepackage{nicefrac}       
\usepackage{microtype}      
\usepackage{xcolor}         

\usepackage{tabularx}
\usepackage{rotating}
\usepackage{multirow}
\usepackage{graphicx}
\usepackage{optidef}
\usepackage{subcaption}
\usepackage{stfloats}
\DeclareMathOperator*{\argmin}{argmin}
\DeclareMathAlphabet{\mathbfcal}{OMS}{cmsy}{b}{n}
\title{ErA: Error-Aware Deep Unrolling Network for Single Image Defocus Deblurring}

\author{%
  Tu Vo \\
  KC Machine Learning Lab\\
  Seoul, South Korea 06181 \\
  \texttt{tuvv@kc-ml2.com} \\
  \And
  Chan Y. Park \\
  KC Machine Learning Lab\\
  Seoul, South Korea 06181 \\
  \texttt{chan.y.park@kc-ml2.com}
}

\begin{document}

\maketitle

\begin{abstract}
  We introduce ErA (Error-Aware Deep Unrolling Network), an end-to-end framework for single-image defocus deblurring. ErA jointly learns a compact kernel basis and per-pixel weights, while an error-aware term in Augmented Lagrangian unrolling corrects kernel estimation errors via alternating updates and ResUNet denoisers. It achieves state-of-the-art PSNR/SSIM on DPDD, RealDOF, and RTF, and shows strong generalization on CUHK without ground truth.
\end{abstract}

\section{Introduction}

Defocus blur arises from circles of confusion caused by camera optics, and its spatially varying size and shape make single-image defocus deblurring especially challenging. Traditional two-stage methods estimate defocus maps and point spread functions (PSF)s before non-blind deblurring \citep{shi2015just, park2017unified, karaali2017edge, lee2019deep, karaali2022svbr}, but errors in PSF estimation often accumulate, leading to artifacts \citep{ji2011robust}. Simplified kernel assumptions, such as disc or Gaussian models \cite{punnappurath2020modeling, quan2021gaussian}, further restrict performance on real-world images. While deep learning has advanced image restoration—most notably for motion deblurring \cite{nah2017deep, tao2018scale, zhang2018dynamic, gao2019dynamic}—directly adapting these models to defocus blur is not effective, since defocus kernels differ significantly from motion kernels. Unrolling-based strategies \cite{quan2021gaussian, quanneumann2023} attempt to model spatially varying blur, but often rely on restrictive Gaussian mixture assumptions.

We present ErA (Error-Aware Unrolling Network), a framework for blind, spatially varying defocus deblurring. ErA explicitly predicts PSFs and embeds error-aware regularization into an Augmented Lagrangian unrolling scheme, alternating closed-form optimization updates with CNN refinements. This hybrid design allows ErA to handle arbitrary, non-Gaussian blur while compensating for kernel estimation errors.

In experiments, ErA achieves state-of-the-art results on DPDD, RealDOF, and RTF benchmarks, and shows strong generalization on CUHK despite the absence of ground truth. These results highlight the potential of error-aware unrolling for robust defocus deblurring in real-world conditions.

\section{Background}

\subsection{Single Image Defocus Deblurring (SIDD)}
A common SIDD pipeline uses Defocus Map Estimation \cite{shi2015just, xu2017estimating, karaali2017edge} followed by Non‐Blind Deblurring \cite{krishnan2009fast}, but this two‐step approach is fragile to estimation errors and assumes simple Gaussian/disc PSFs. Deep end‐to‐end methods overcame this—e.g., Abuolaim and Brown’s CNN \cite{abuolaim2020defocus}, per‐pixel filter prediction \cite{lee2021iterative}, dynamic residual blocks \cite{ruan2022learning}, Son \textit{et al.}’s kernel‐sharing atrous convolutions \cite{son2021single}, and Quan \textit{et al.}’s duplex scale‐recurrent network \cite{quanneumann2023}. More recent transformer‐based designs (MultiPyramid \cite{zhang2024unified}, inverse‐kernel \cite{aaai2024_inverse_kernel}) and advanced architectures (IRNeXt \cite{cui2023irnext}, Selective Frequency \cite{cui2023selective}, Frequency Selection \cite{cui2023image}, SSMNet \cite{gao2024learning}, NeumannNet \cite{quanneumann2023}) improve restoration but still assume parametric or uniform PSFs. These assumptions limit their generalization to complex, real-world optics. In contrast, ErA learns a compact PSF basis and dense per‐pixel weights to synthesize fully non‐uniform, non‐Gaussian blur kernels, enabling robust defocus removal across diverse blur patterns.

\begin{figure}[t]
    \centering
    \includegraphics[width=0.45\textwidth]{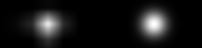}
    \caption{The predicted defocus blur kernel (\textit{left}) vs widely assumed Gaussian kernel. (\textit{right}).}
    \label{fig:showcase}
\end{figure}

\subsection{Unrolling-Based Deep Neural Networks}
Unrolling-based deep neural networks have been widely adopted for image deblurring, particularly in non-blind settings where the PSF is known \cite{zhang2017learning, dong2018denoising, nan2020variational}. In parallel, several approaches have explored blind image deblurring using unrolling techniques, where the PSF is not provided \cite{schuler2015learning, zuo2016learning, li2019algorithm}. However, these methods generally assume uniform or parameterized blur kernels, limiting their applicability. More recently, Quan et al. \cite{quan2021gaussian} applied unrolling to defocus deblurring under the assumption of a Gaussian-shaped kernel, which restricts the model's ability to handle arbitrary PSFs. Vo et al. \cite{vo2025deep} further extended this direction by addressing spatially varying blur kernels, but their approach does not incorporate any mechanism for error correction.

Recent work on Adaptive Basis Decomposition \cite{carbajal2021non} proposes estimating a set of global blur bases and a per-pixel weight map to model spatially varying blur. However, it does not incorporate any iterative reconstruction or unrolling, nor does it explicitly address kernel estimation errors. Separately, Deep Learning for Handling Kernel/Model Uncertainty \cite{nan2020deep} introduces a sparse error-correction term in uniform deblurring, but without supporting spatial variation. ErA combines both ideas: it adopts a basis+weight representation for PSFs and embeds an error-aware term into an unrolled ALM framework, enabling robust restoration from arbitrary, non-Gaussian defocus blur.

\section{Proposed Method}

\subsection{Problem Formulation}
We model blind deblurring as
\begin{mini}|l|
{\mathbfcal{H}, \mathbfcal{X}}{\tfrac{1}{2}\|\mathbfcal{H}\otimes \mathbfcal{X} - \mathbfcal{Y} \|_{2}^{2}}
{}{}
\label{eq:original}
\end{mini}
where \(\mathbfcal{H}\), \(\mathbfcal{X}\), and \(\mathbfcal{Y}\) denote blur kernel, clean image, and observed blur, respectively. Since kernel estimates are imperfect, we introduce an error term \(\mathcal{E}\), assumed sparse:
\begin{mini}|l|
{\mathbfcal{X},\mathbfcal{H},\mathcal{E}}{\tfrac{1}{2}\|\mathbfcal{H}\otimes \mathbfcal{X} - \mathbfcal{Y} + \mathcal{E}\|_{2}^{2} + \|\mathcal{E}\|_{1}}
{}{}
\label{eq:originalwithe}
\end{mini}

To improve generalization, we add regularizers \(\phi(\mathbfcal{X})\) and \(f(\mathcal{E})\), yielding:
\begin{mini}|l|
{\mathbfcal{X},\mathbfcal{H},\mathcal{E}}{\tfrac{1}{2}\|\mathbfcal{H}\otimes \mathbfcal{X} - \mathbfcal{Y} + \mathcal{E}\|_{2}^{2} + \|\mathcal{E}\|_{1} + \phi(\mathbfcal{X}) + f(\mathcal{E})}
{}{}
\label{eq:withmodels}
\end{mini}

\subsection{Optimization}\label{sec:opt}
We solve \eqref{eq:withmodels} via the Augmented Lagrangian Method (ALM) with variable splitting, iteratively and individually.
\begin{mini}|l|
{\mathbfcal{X},\mathbfcal{H},\mathcal{E},\mathbfcal{U},\mathbfcal{P},\mathbfcal{Z}}{\tfrac{1}{2}\|\mathbfcal{U} - \mathbfcal{Y}+\mathcal{E}\|_{2}^{2} + \phi(\mathbfcal{Z}) + \lambda_{3}\|\mathcal{E}\|_{1} + f(\mathbfcal{P})}
{}{}
\addConstraint{\mathbfcal{U}=\mathbfcal{H}\otimes \mathbfcal{X},\ \mathbfcal{P}=\mathcal{E},\ \mathbfcal{Z}=\mathbfcal{X}}
\label{eq:withauxies}
\end{mini}

ALM yields closed-form updates for \(\mathbfcal{U}, \mathbfcal{X}\), soft-threshold\citep{hale2008fixed} for \(\mathbfcal{E}\), while \(\mathbfcal{Z}, \mathbfcal{P}\) are updated via CNN-based operators \(\mathbfcal{D}_\phi\) and \(\mathbfcal{D}_f\) (ResUNet \citep{diakogiannis2020resunet}). Multiplier updates follow standard ALM steps. Please refer to the \hyperref[sec:appendix]{Appendix} section below for more details on solving.


\begin{equation}
   \mathbfcal{U}_{t+1} = \argmin_{\mathbfcal{U}} f(\mathbfcal{U}) = \frac{\lambda_{1}\mathbfcal{H}\otimes\mathbfcal{X}_{t} + \mathbf{\Gamma}_{t} + \mathbfcal{Y} - \mathbfcal{E}_{t}}{1+\lambda_{1}}
\end{equation}


\begin{equation}
       \mathbfcal{X}_{t+1} = \argmin_{\mathbfcal{X}} f(\mathbfcal{X}) = \mathbfcal{F}^{-1}\left\{\frac{\mathbfcal{F}(\mathbfcal{H}^{T}(-\mathbf{\Gamma}_{t} + \lambda_{1}\mathbfcal{U}_{t+1}) - \mathbf{\Omega}_{t} + \lambda_{2}\mathbfcal{Z}_{t+1})}{\lambda_{1}\mathbfcal{F}(\mathbfcal{H})^{2} + \lambda_{2}}\right\}
\end{equation} 

\begin{equation}
   \mathbfcal{E}_{t+1}=\argmin_{\mathbfcal{E}}f(\mathbfcal{E}) = soft-thresh(-\frac{\mathbf{\Delta}_{t} + \mathbfcal{U}_{t+1}-\mathbfcal{Y} - \lambda_{3} \mathbfcal{P}_{t}}{\lambda_{3} + 1})
   \label{eq:et13}
\end{equation} 

\begin{equation}
 \mathbfcal{Z}_{t+1} = \argmin_{\mathbfcal{Z}} f(\mathbfcal{Z}) = \mathbfcal{D}_{\phi}(\mathbfcal{X}_{t} + \frac{\mathbf{\Omega}_{t}}{\lambda_{2}}), \mathbfcal{P}_{t+1} = \argmin_{\mathbfcal{P}} f(\mathbfcal{P}) = \mathbfcal{D}_{f}(\mathbfcal{P} - \frac{\mathbf{\Delta_{t}} + \lambda_{3}\mathbfcal{E}_{t+1}}{\lambda_{3}})
   \label{eq:pt1} 
\end{equation} 

\begin{equation}
\begin{aligned} 
   \mathbf{\Gamma}_{t+1} = \mathbf{\Gamma}_{t} + \lambda_{1} * (\mathbfcal{H}\otimes\mathbfcal{X}_{t+1} - \mathbfcal{U}_{t+1}) \\
   \mathbf{\Omega}_{t+1} = \mathbf{\Omega}_{t} + \lambda_{2} * (\mathbfcal{X}_{t+1} - \mathbfcal{Z}_{t+1}) \\
   \mathbf{\Delta}_{t+1} = \mathbf{\Delta}_{t} + \lambda_{3} * (\mathbfcal{E}_{t+1} - \mathbfcal{P}_{t+1}) 
\end{aligned}
\end{equation}

\subsection{Network Architecture}
\begin{figure*}[t]
    \centering
    \includegraphics[width=0.95\textwidth]{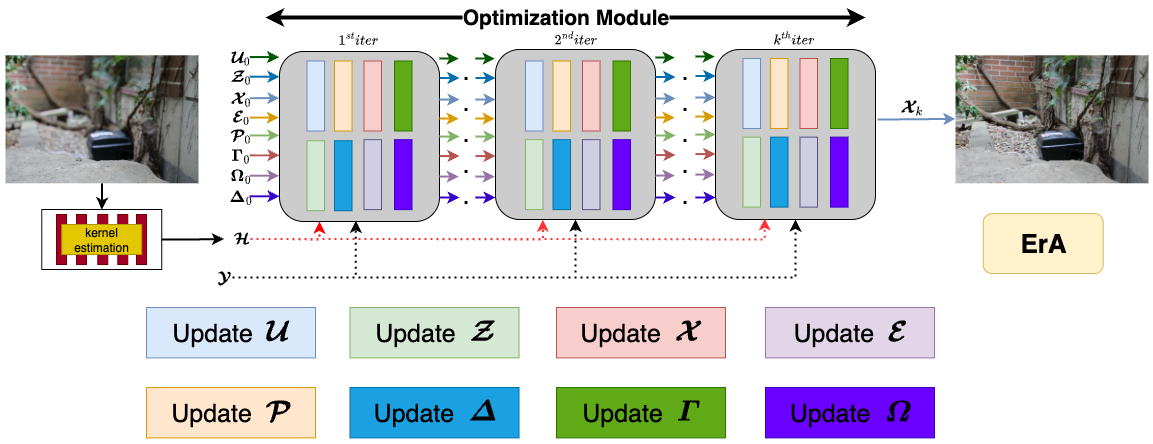}
    \caption{ErA architecture. $\mathbfcal{X}_{0}$ is initialized by convolving $\mathbfcal{Y}$ with $\mathbfcal{H}$. Each of the $K$ blocks updates $\mathbfcal{U}$, $\mathbfcal{E}$, and $\mathbfcal{X}$ in closed form, while $\mathbfcal{Z}$ and $\mathbfcal{P}$ use a ResUNet CNN.}
    \label{fig:edunet}
\end{figure*}

We unroll $K$ iterations of ALM into a deep network, \textbf{ErA} (Error-Aware Unrolling). As shown in Fig.~\ref{fig:edunet}, each block updates \(\mathbfcal{U}, \mathbfcal{E}, \mathbfcal{X}\) in closed form, while CNN modules refine \(\mathbfcal{Z}\) and \(\mathbfcal{P}\). A kernel estimation CNN predicts a global kernel and spatial weight map, enabling pixel-wise PSF modeling.

\subsection{Training Loss}
The network is trained with a combined loss \(L = \omega \|\mathbfcal{X}_{\text{pred}} - \mathbfcal{X}_{\text{gt}}\|_1 
    + (1-\omega)\|\mathbfcal{H}\otimes \mathbfcal{X}_{\text{pred}} - \mathbfcal{Y}\|_1\) balancing image reconstruction and kernel consistency.

\section{Experiments}  
\subsection{Benchmark and Implementation Details}  
\subsubsection{Benchmark Datasets}  
We train and evaluate our method on the DPDD dataset \cite{abuolaim2020defocus}, which contains 500 dual-pixel image pairs in 16-bit format, split into 350 training, 74 validation, and 76 test samples. Additional evaluation is performed on the RealDOF \cite{lee2021iterative} and RTF \cite{d2016non} datasets.

\subsubsection{Implementation Details}  
We set the unrolling depth to \(K=10\) and the blur kernel size to \(61 \times 61\). Training is performed end-to-end using Adam \cite{kingma2014adam} for 150 epochs. Training samples are cropped into \(140 \times 140\) patches with flips/rotations for augmentation. Implementation is in PyTorch on an NVIDIA RTX A6000.  

\begin{table*}[th!]
\addtolength{\tabcolsep}{-0.25em}
\centering
\begin{tabular}{c|c|c|c|c|c|c|c|c|c}
\hline
\multirow{2}{*}{Model} & \multicolumn{3}{c|}{DPDD} & \multicolumn{3}{c|}{RealDOF} & \multicolumn{3}{c}{RTF} \\
\cline{2-10}
 & PSNR $\uparrow$ & SSIM $\uparrow$ & LPIPS $\downarrow$ & PSNR $\uparrow$ & SSIM $\uparrow$ & LPIPS $\downarrow$ & PSNR $\uparrow$ & SSIM $\uparrow$ & LPIPS $\downarrow$ \\
\hline
DRBNet       & 25.485 & 0.792 & 0.254 & 24.700 & 0.744 & 0.337 & 24.463 & 0.773 & 0.311 \\
NRKNet       & 26.110 & 0.810 & 0.220 & 25.060 & \underline{0.767} & 0.339 & \textbf{25.931} & \underline{0.829} & 0.216 \\
P2IKT        & 26.280 & 0.807 & \textbf{0.191} & 25.480 & 0.762 & \underline{0.306} & 25.260 & 0.819 & \textbf{0.207} \\
IRNeXT       & 26.300 & \underline{0.814} & \underline{0.206} & \underline{25.660} & 0.755 & 0.336 & 25.333 & \textbf{0.854} & 0.249 \\
ErA w/o \(\mathbfcal{E}\) & \underline{26.361} & 0.812 & 0.226 & 25.422 & 0.764 & 0.330 & 25.084 & 0.820 & 0.243 \\
\textbf{ErA} & \textbf{26.687} & \textbf{0.815} & 0.219 & \textbf{25.747} & \textbf{0.772} & 0.319 & \underline{25.502} & 0.823 & \underline{0.215} \\
\hline
\end{tabular}
\caption{Quantitative results on DPDD, RealDOF, and RTF. Best and second-best are highlighted in bold/underlined. With error constraint, ErA outperforms state-of-the-art models.}
\label{tab:table1}
\end{table*}

\subsection{Performance Comparison}  
We benchmark ErA against some NN-based methods: DRBNet \citep{ruan2022learning}, NRKNet \citep{quanneumann2023}, P2IKT \citep{aaai2024_inverse_kernel}, and IRNeXt \cite{cui2023irnext}. Results are either reported from the original papers or reproduced with public models. 

\subsubsection{Quantitative Evaluation}  
Table~\ref{tab:table1} shows that ErA achieves the highest PSNR and SSIM on both DPDD and RealDOF, and competitive results on RTF. Despite slightly higher LPIPS than some baselines, our method yields structurally faithful reconstructions (e.g., \textbf{+0.4 dB on DPDD}, \textbf{+0.1 dB on RealDOF}).

\subsubsection{Qualitative Evaluation}  
Figures~\ref{fig:ddp} illustrates that ErA recovers sharper text, wire edges, and fine structures compared to competing methods. The explicit error-aware mechanism reduces artifacts and enhances sharpness. 

\begin{figure*}[h]
\centering

\begin{subfigure}{0.16\textwidth}
    \centering \small \textbf{Input}
\end{subfigure}%
\begin{subfigure}{0.16\textwidth}
    \centering \small \textbf{IRNext}
\end{subfigure}%
\begin{subfigure}{0.16\textwidth}
    \centering \small \textbf{NRKNet}
\end{subfigure}%
\begin{subfigure}{0.16\textwidth}
    \centering \small \textbf{P2IKT}
\end{subfigure}%
\begin{subfigure}{0.16\textwidth}
    \centering \small \textbf{ErA}
\end{subfigure}%
\begin{subfigure}{0.16\textwidth}
    \centering \small \textbf{GT}
\end{subfigure}
\begin{subfigure}{0.16\textwidth}
    \includegraphics[width=\linewidth]{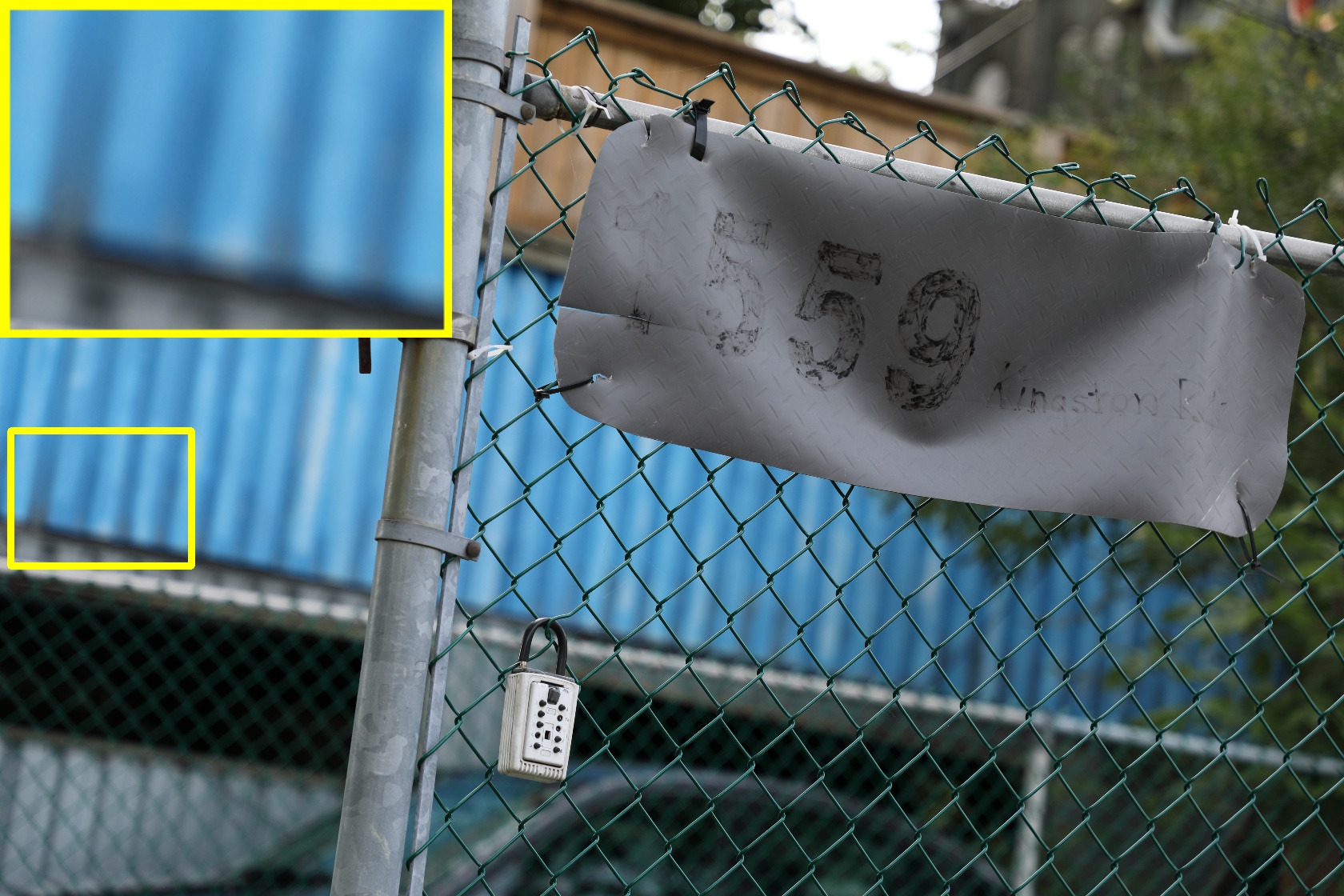}
\end{subfigure}%
\begin{subfigure}{0.16\textwidth}
    \includegraphics[width=\linewidth]{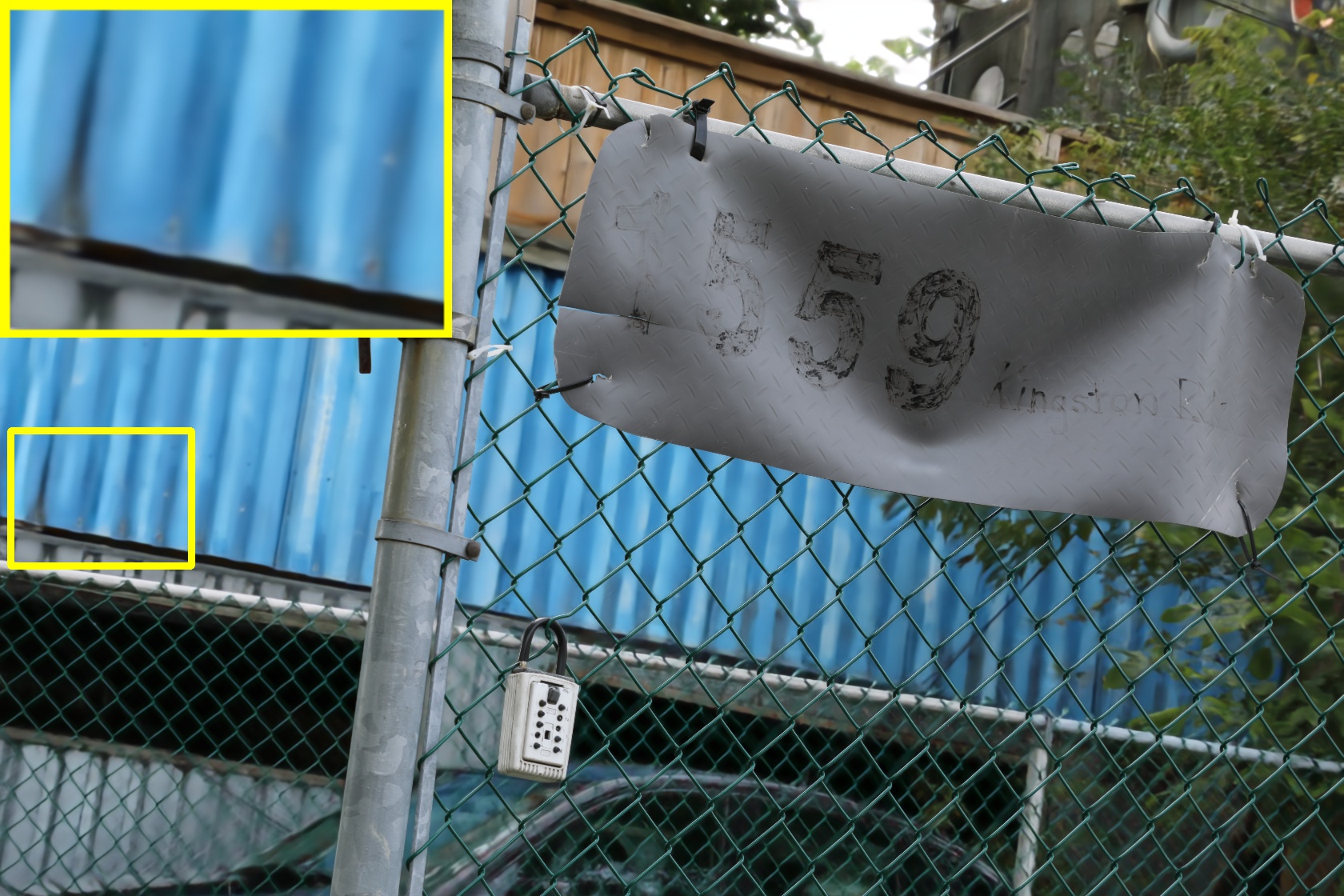}
\end{subfigure}%
\begin{subfigure}{0.16\textwidth}
    \includegraphics[width=\linewidth]{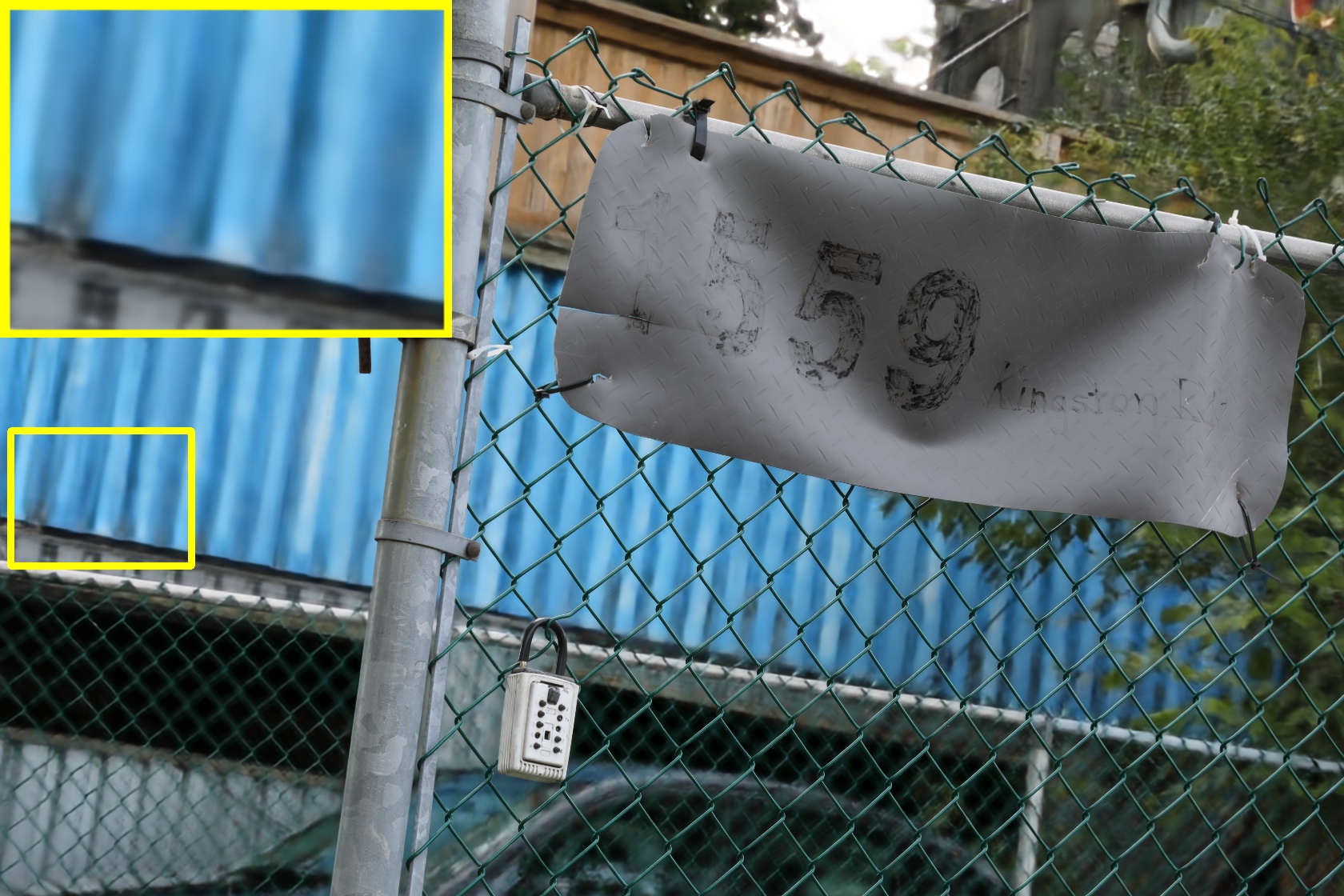}
\end{subfigure}%
\begin{subfigure}{0.16\textwidth}
    \includegraphics[width=\linewidth]{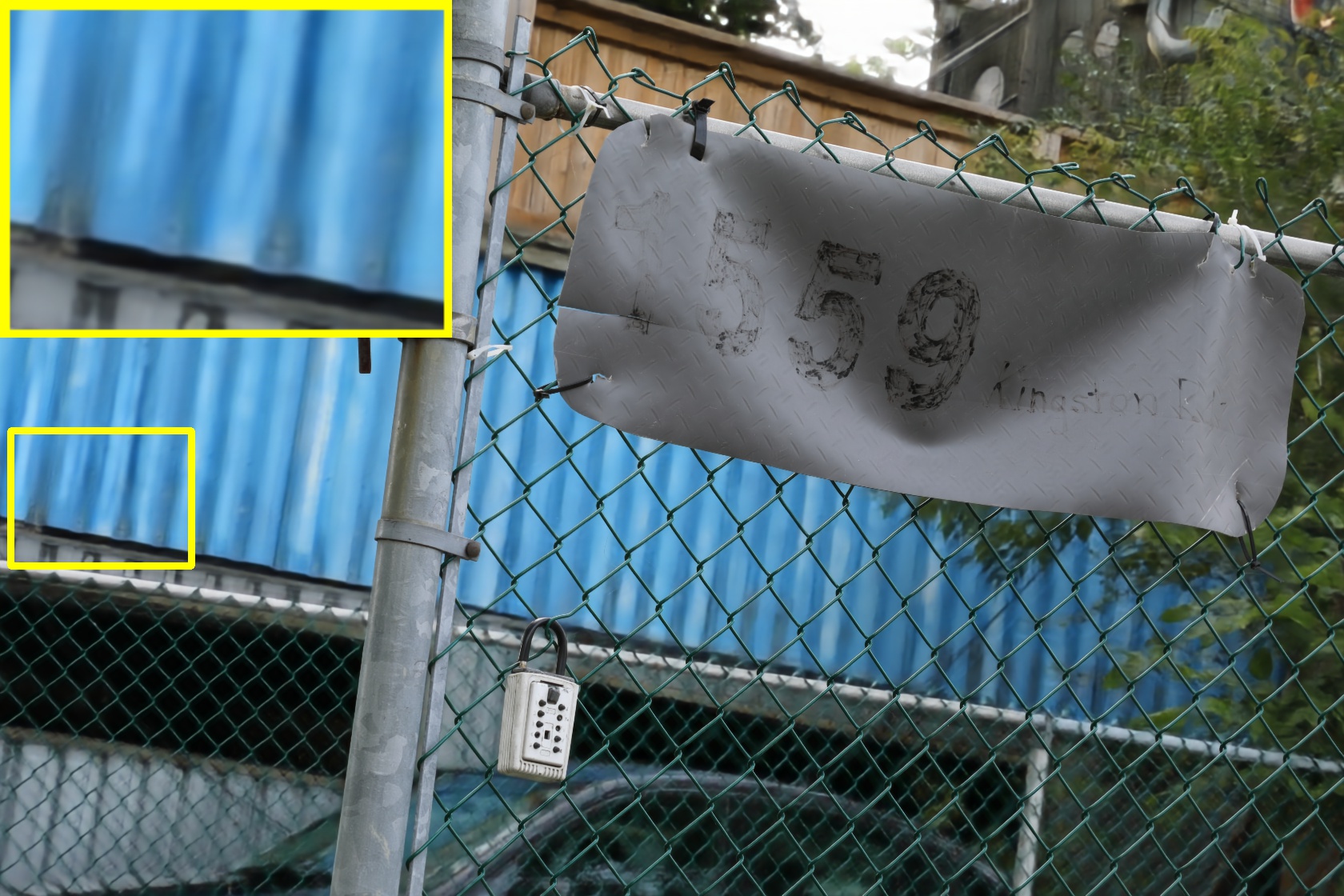}
\end{subfigure}%
\begin{subfigure}{0.16\textwidth}
    \includegraphics[width=\linewidth]{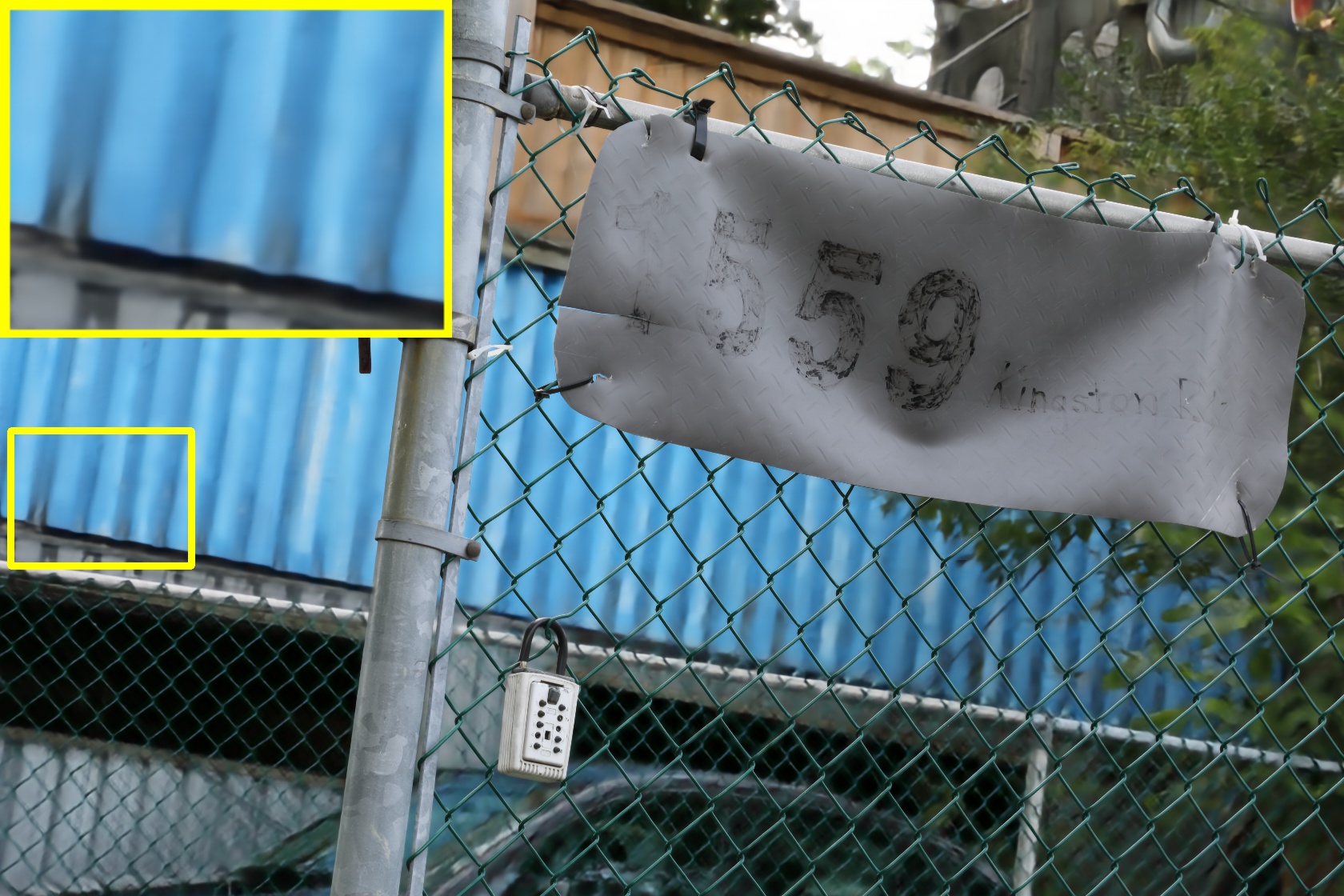}
\end{subfigure}%
\begin{subfigure}{0.16\textwidth}
    \includegraphics[width=\linewidth]{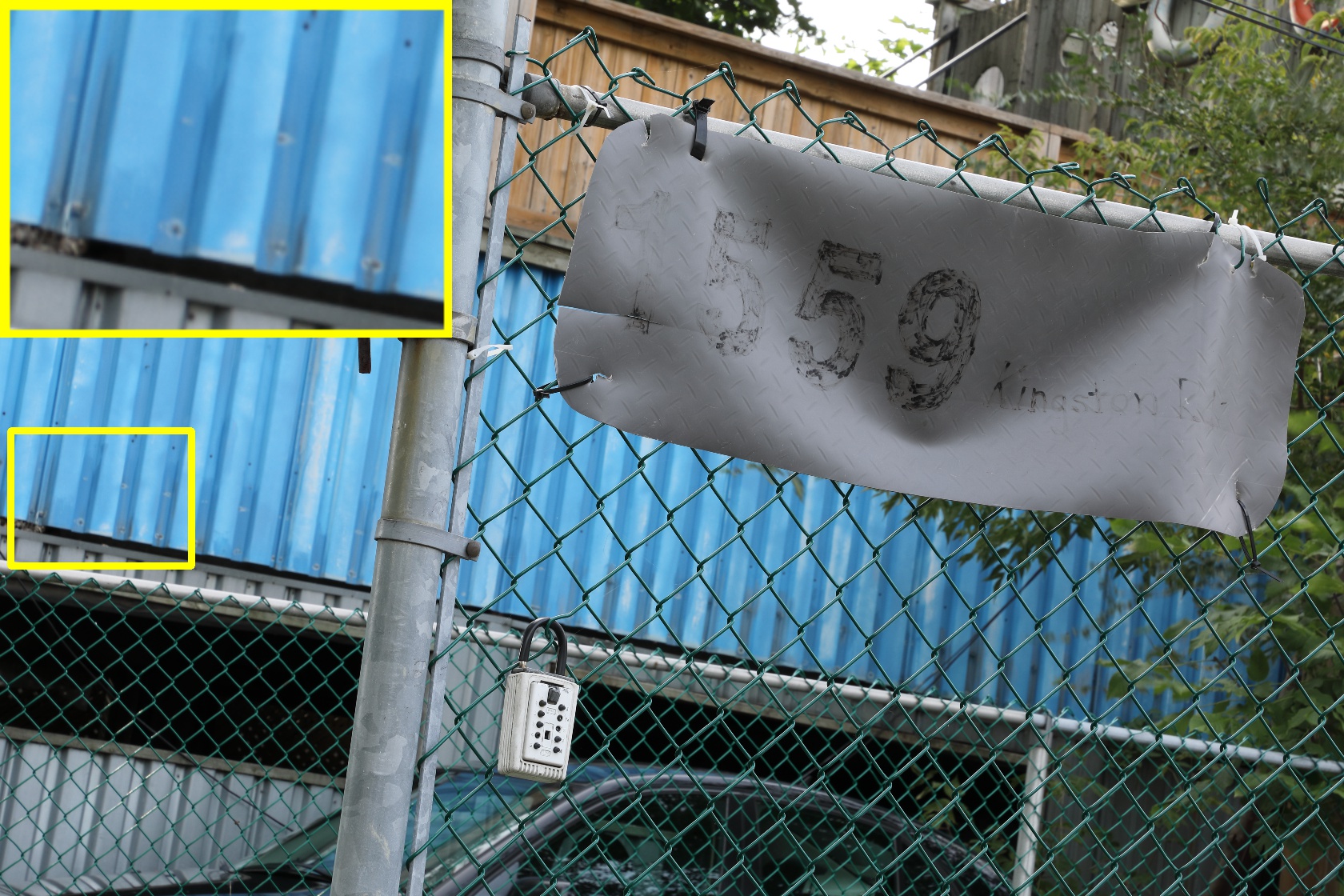}
\end{subfigure}

\vspace{0.5mm}

\begin{subfigure}{0.16\textwidth}
    \includegraphics[width=\linewidth]{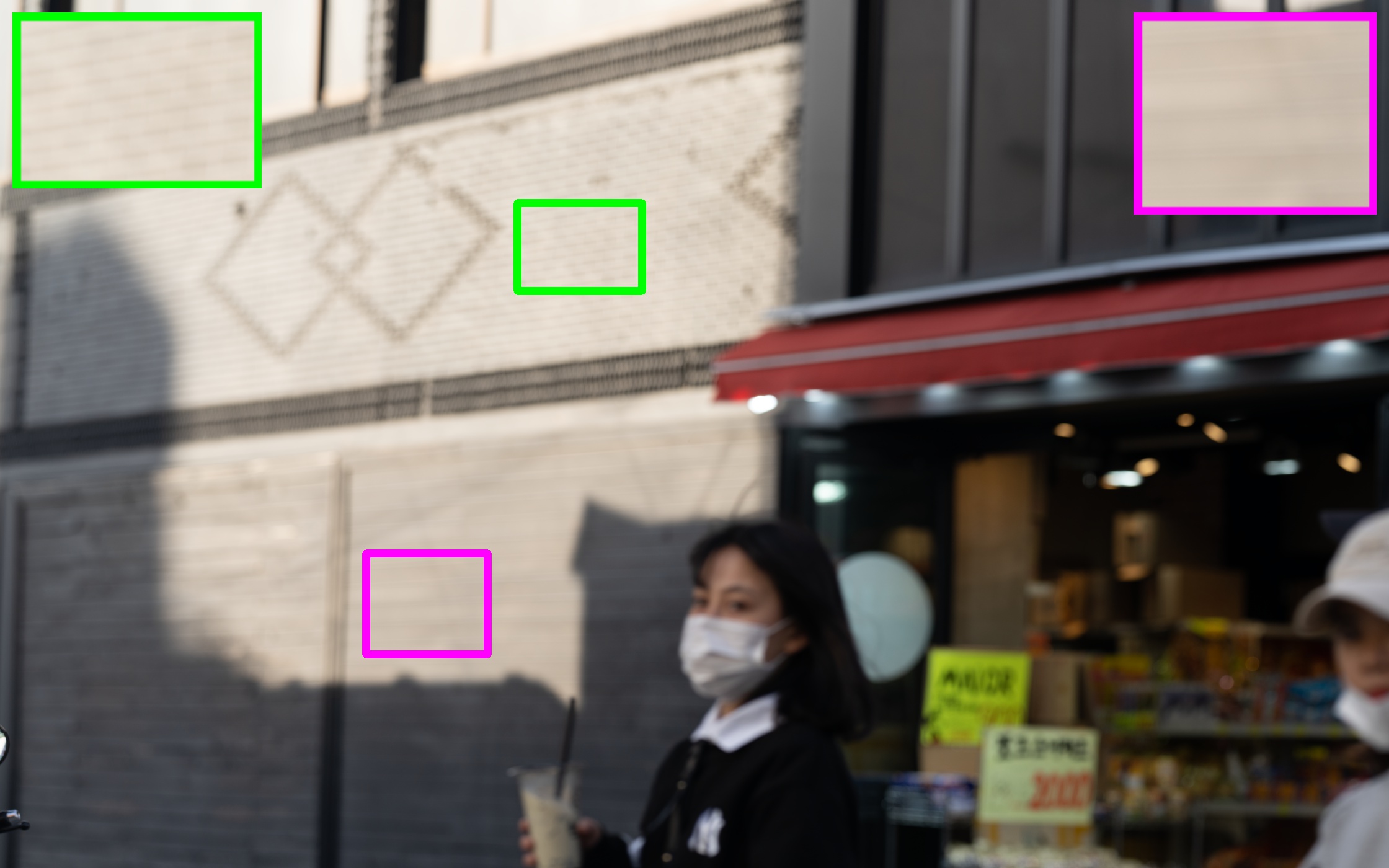}
\end{subfigure}%
\begin{subfigure}{0.16\textwidth}
    \includegraphics[width=\linewidth]{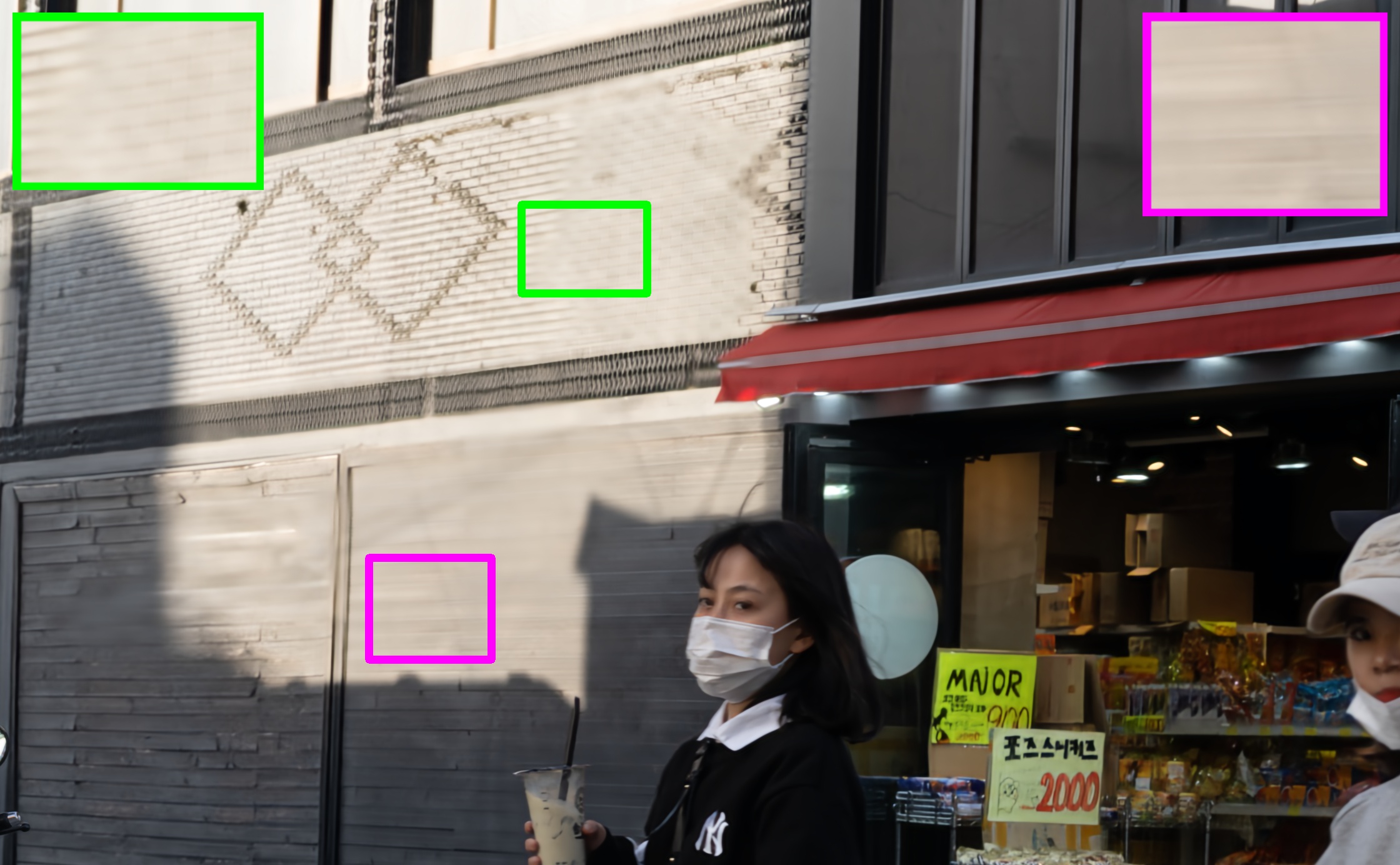}
\end{subfigure}%
\begin{subfigure}{0.16\textwidth}
    \includegraphics[width=\linewidth]{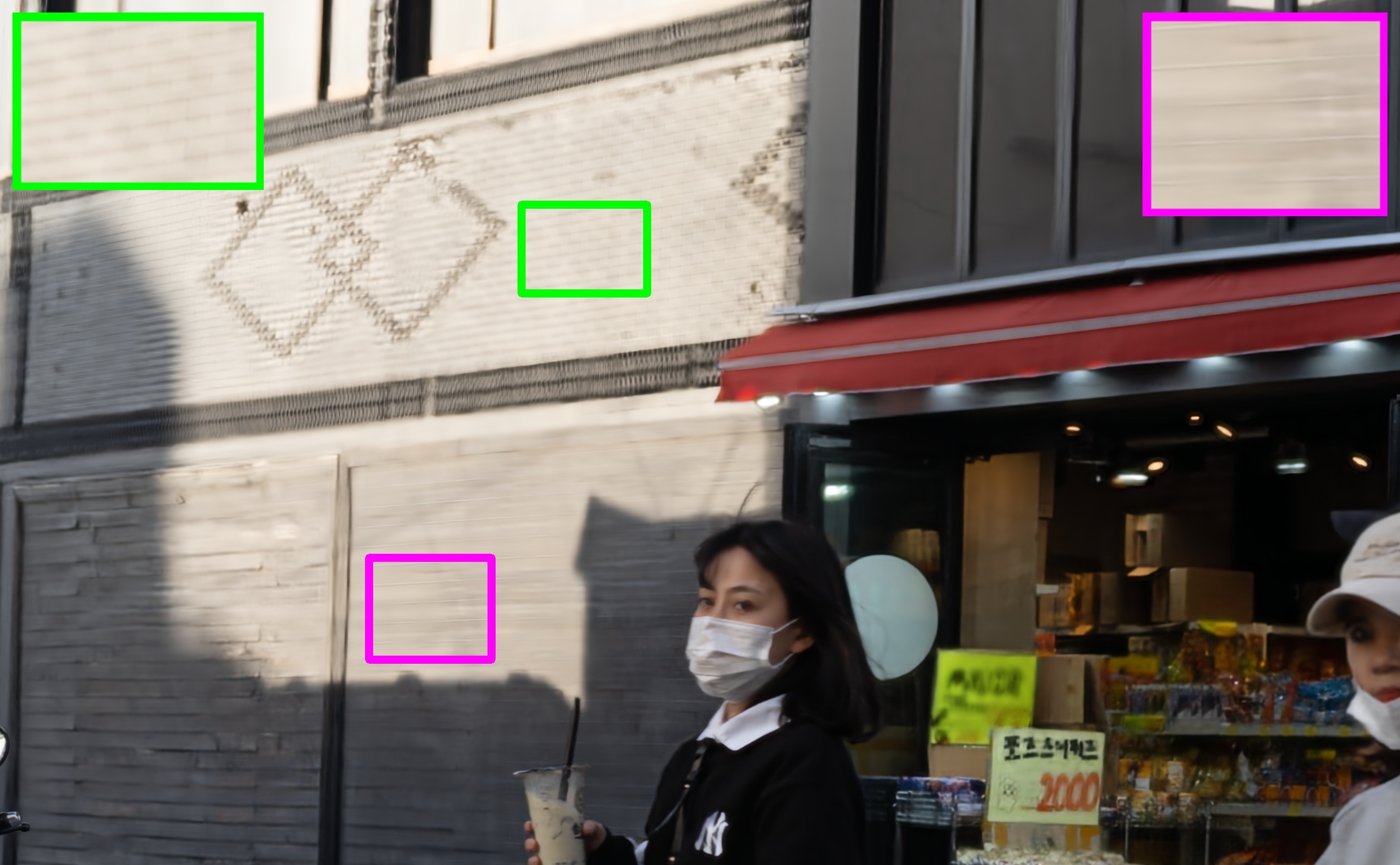}
\end{subfigure}%
\begin{subfigure}{0.16\textwidth}
    \includegraphics[width=\linewidth]{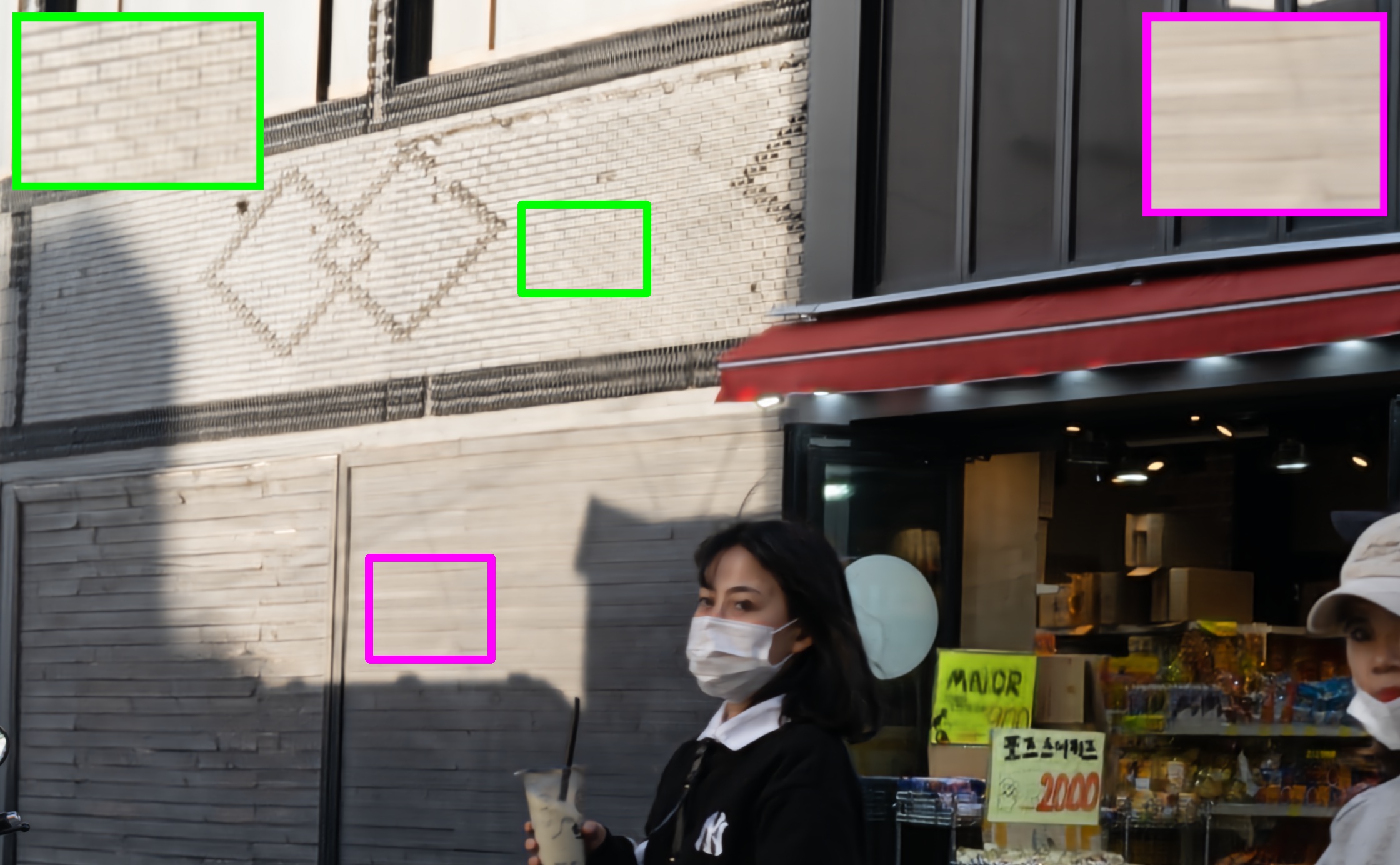}
\end{subfigure}%
\begin{subfigure}{0.16\textwidth}
    \includegraphics[width=\linewidth]{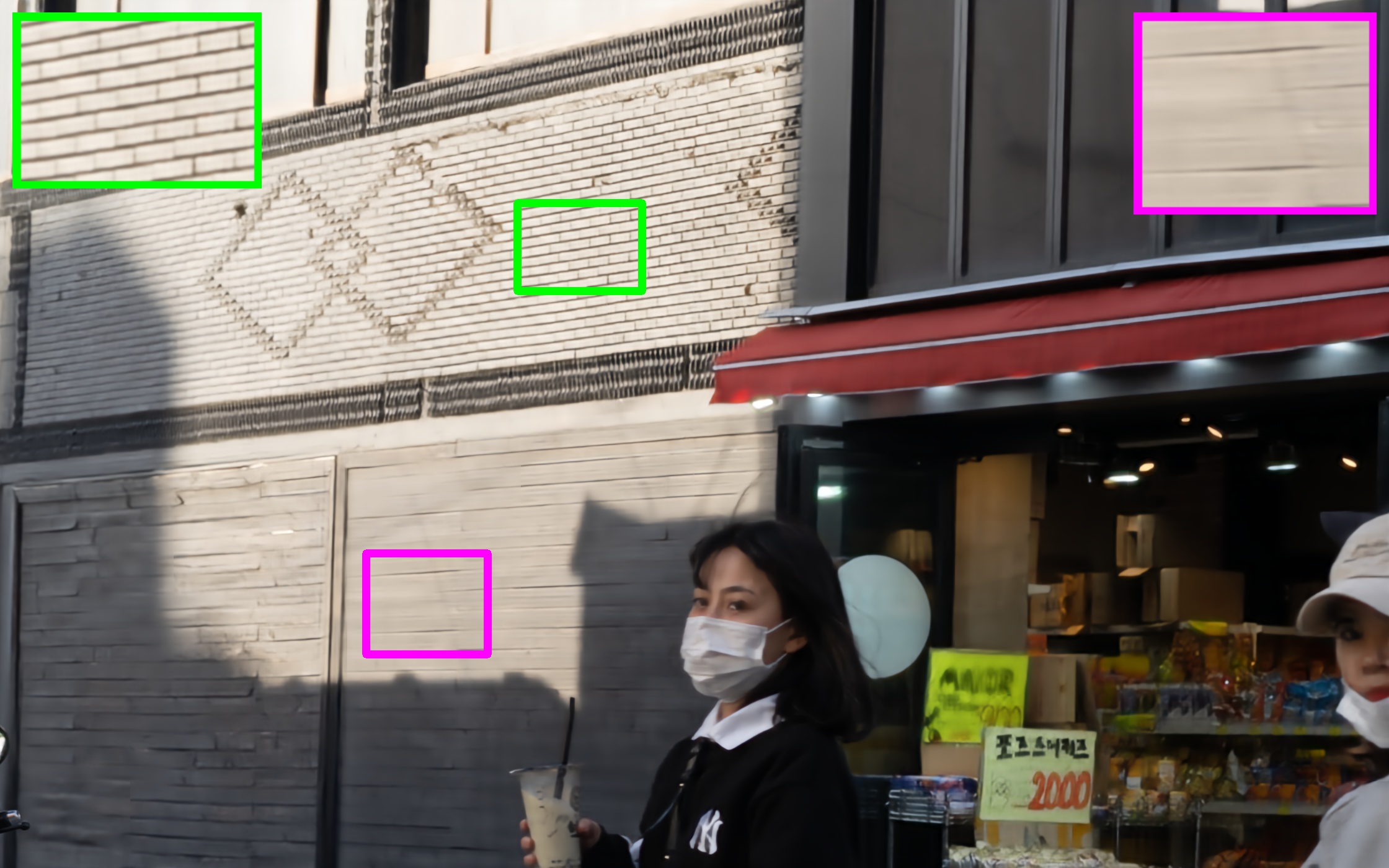}
\end{subfigure}%
\begin{subfigure}{0.16\textwidth}
    \includegraphics[width=\linewidth]{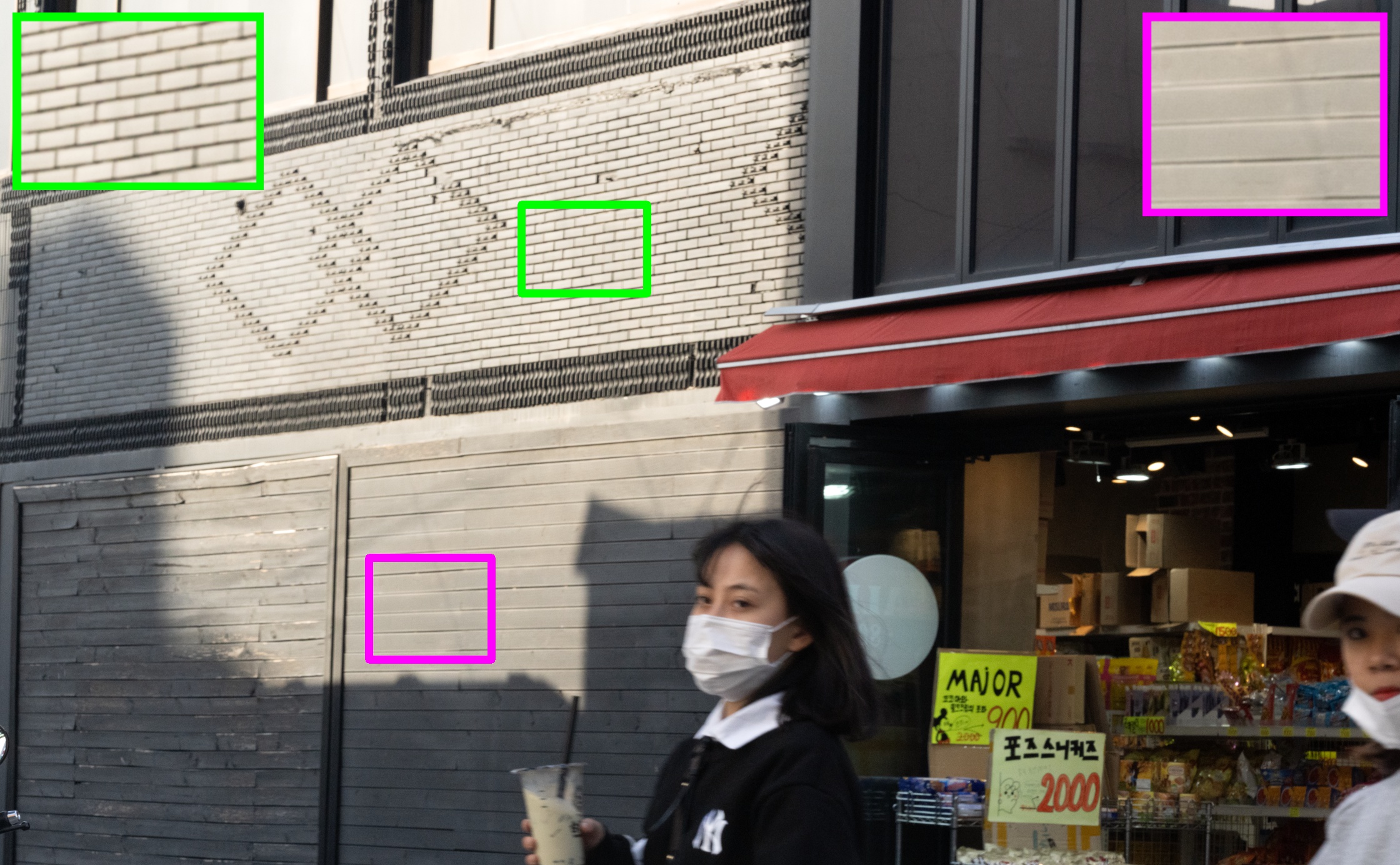}
\end{subfigure}

\caption{Visual Comparison of various methods on DPDD (top) and RealDOF (bottom) test set.}
\label{fig:ddp}
\end{figure*}

\begin{figure*}[h]
\centering

\begin{subfigure}{0.16\textwidth}
    \centering \small \textbf{Input}
\end{subfigure}%
\begin{subfigure}{0.16\textwidth}
    \centering \small \textbf{IRNext}
\end{subfigure}%
\begin{subfigure}{0.16\textwidth}
    \centering \small \textbf{NRKNet}
\end{subfigure}%
\begin{subfigure}{0.16\textwidth}
    \centering \small \textbf{P2IKT}
\end{subfigure}%
\begin{subfigure}{0.16\textwidth}
    \centering \small \textbf{ErA}
\end{subfigure}%

\begin{subfigure}{0.18\textwidth}
    \includegraphics[width=\linewidth]{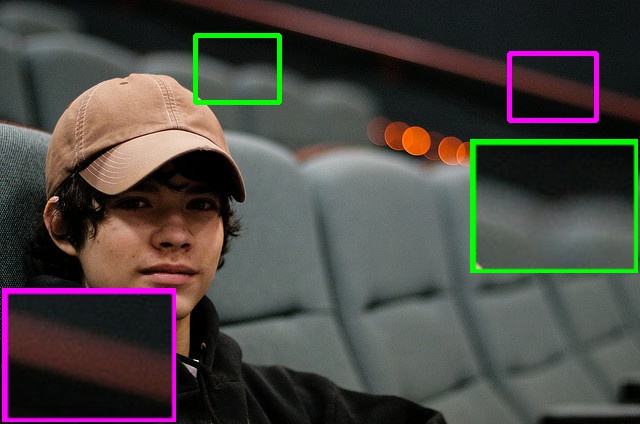}
\end{subfigure}%
\begin{subfigure}{0.18\textwidth}
    \includegraphics[width=\linewidth]{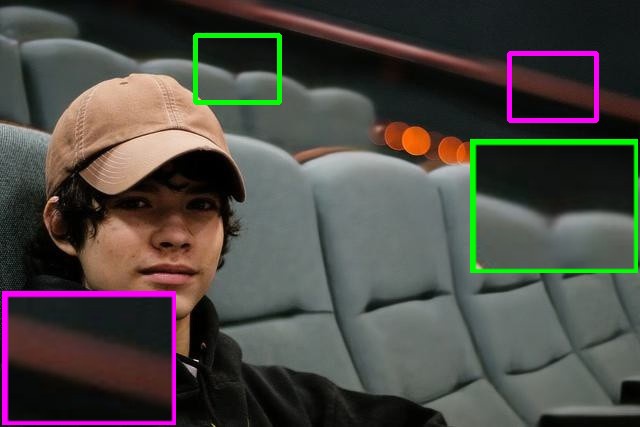}
\end{subfigure}%
\begin{subfigure}{0.18\textwidth}
    \includegraphics[width=\linewidth]{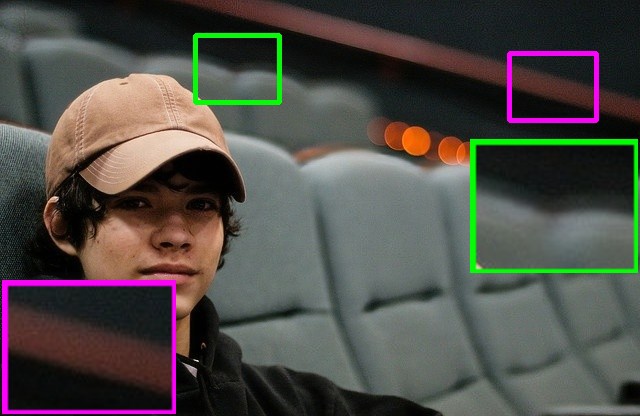}
\end{subfigure}%
\begin{subfigure}{0.18\textwidth}
    \includegraphics[width=\linewidth]{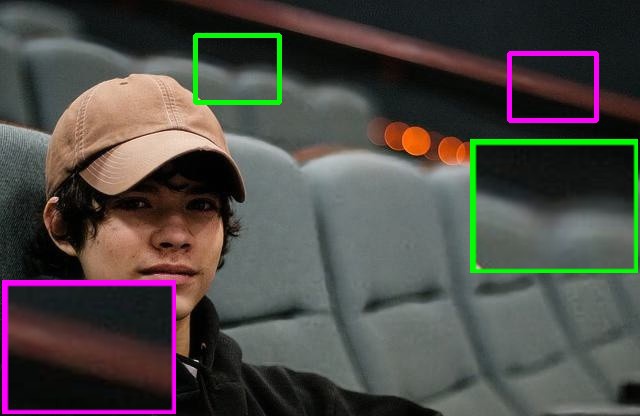}
\end{subfigure}%
\begin{subfigure}{0.18\textwidth}
    \includegraphics[width=\linewidth]{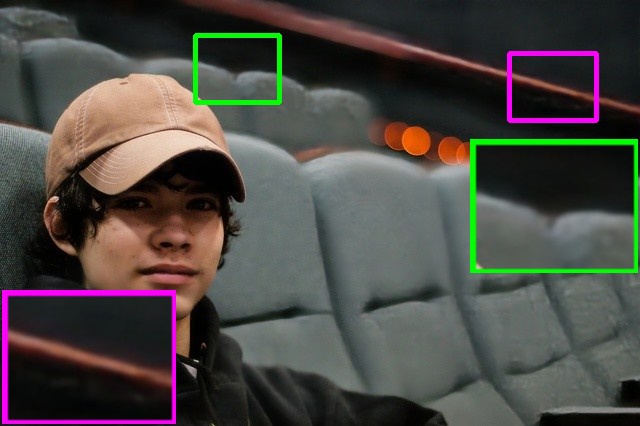}
\end{subfigure}%

\caption{Visual Comparison of various methods CUHK test set.}
\label{fig:ddp}
\end{figure*}

\section{Conclusion}  
We propose a deep unrolling network with an error-aware module for single-image defocus deblurring. By treating it as a blind, non-uniform blur problem, our model iteratively solves the optimization via ALM with closed-form updates and denoisers. Experiments show that ErA outperforms state-of-the-art methods with fewer training samples and better generalization.

\bibliographystyle{abbrvnat} 
\bibliography{coml_2025}
\clearpage
\section{Appendix}\label{sec:appendix}
\subsection{Derivation of closed-form solution}
In this section, we will provide the complete closed-form solutions for each variable described in \textbf{Section \ref{sec:opt}} above.

\textbf{Update} \(\mathbfcal{U}_{t+1}\): At the \textit{t}th iteration, we update \(\mathbfcal{U}\) as
\nolinebreak
\begin{equation}
\begin{aligned}
    \mathbfcal{U}_{t+1} = \argmin_{\mathbfcal{U}}\frac{1}{2}\|\mathbfcal{U} + \mathcal{E}_{t} - \mathbfcal{Y}\|_{2}^{2} + \langle\mathbf{\Gamma}_{t}, \mathbfcal{H}\otimes\mathbfcal{X}_{t}-\mathbfcal{U}\rangle \\
    + \frac{\lambda_{1}}{2}\|\mathbfcal{H}\otimes\mathbfcal{X}_{t}-\mathbfcal{U}\|_{2}^{2}
    \label{eq:updateU}
\end{aligned}
\end{equation} 

Taking the derivative over \(\mathbfcal{U}\) and set to \(0\), we have:
\begin{equation}
   \mathbfcal{U}_{t} +\mathbfcal{E}_{t} - \mathbfcal{Y} - \mathbf{\Gamma}_{t}   + \lambda_{1} (\mathbfcal{U}_{t} - \mathbfcal{H}\otimes\mathbfcal{X}_{t}) = 0
\end{equation}
\begin{equation}
   \Leftrightarrow \mathbfcal{U}_{t}(1 + \lambda_{1}) = -\mathbfcal{E}_{t} + \mathbfcal{Y} + \mathbf{\Gamma}_{t}  - \lambda_{1} \mathbfcal{H}\otimes\mathbfcal{X}_{t}
\end{equation}
From here, \(\mathbfcal{U}_{t+1}\) can be calculated as follow:
\begin{equation}
   \mathbfcal{U}_{t+1} = \frac{\lambda_{1}\mathbfcal{H}\otimes\mathbfcal{X}_{t} + \mathbf{\Gamma}_{t} + \mathbfcal{Y} - \mathbfcal{E}_{t}}{1+\lambda_{1}}
\end{equation}

\textbf{Update} \(\mathbfcal{E}_{t+1}\):
\nolinebreak
\begin{equation}
\begin{aligned}
    \mathbfcal{E}_{t+1}=\argmin_{\mathbfcal{E}}\frac{1}{2}\|\mathbfcal{U}_{t+1} -  \mathbfcal{Y} + \mathbfcal{E}\|_{F}^2 + \lambda_{3}\|\mathbfcal{E}\|_{1} \\
    + \langle\mathbf{\Delta}_{t}, \mathbfcal{E} - \mathbfcal{P}_{t}\rangle + \frac{\lambda_{3}}{2}\|\mathbfcal{E} - \mathbfcal{P}_{t}\|_{F}^{2}
\end{aligned}
\end{equation}

\begin{equation}
   =\argmin_{\mathbfcal{E}}\lambda_{3}\|\mathbfcal{E}\|_{1} + (\lambda_{3}+1)\|\mathbfcal{E}\|^{2}_{2}       
   + \langle \mathbfcal{N}, \mathbfcal{E}\rangle
\label{eq:et1}
\end{equation} 
where \(\mathbf{\Delta}_{t} + \mathbfcal{U}_{t+1}-\mathbfcal{Y} - \lambda_{3} \mathbfcal{P}_{t} = \mathbfcal{N}\). Let \(\lambda_{3} + 1 = \mathbf{\Phi}\), (\ref{eq:et1}) become
\begin{equation}
   \mathbfcal{E}_{t+1}=\argmin_{\mathbfcal{E}}\lambda_{3}\|\mathbfcal{E}\|_{1} + \|\sqrt{\Phi}\mathbfcal{E} + \frac{1}{\sqrt{\mathbf{\Phi}}}\mathbfcal{N}\|_{F}^{2}
\end{equation} 

\begin{equation}
   =\argmin_{\mathbfcal{E}}\frac{\lambda_{3}}{\Phi}\|\mathbfcal{E}\|_{1} + \frac{1}{2}\|\mathbfcal{E} - \frac{1}{\mathbf{\Phi}}(-\mathbfcal{N})\|_{F}^{2}
   \label{eq:et13}
\end{equation} 

The closed form solution for Equation (\ref{eq:et13}) can be derived by employing the soft thresholding operator, as outlined in Reference \cite{hale2008fixed}.\\

\textbf{Update} \(\mathbfcal{Z}_{t+1}\):

\begin{equation}
\mathbfcal{Z}_{t+1}=\argmin_{\mathbfcal{Z}}\phi(\mathbfcal{Z}_{t+1}) + \langle\mathbf{\Omega}_{t}, \mathbfcal{X}_{t}-\mathbfcal{Z}_{t}\rangle + \frac{\mathbf{\lambda}_{2}}{2}\|\mathbfcal{X}_{t}-\mathbfcal{Z}_{t}\|_{2}^{2}
\end{equation}


\begin{equation*}
\begin{aligned}[t]
&=\argmin_{\mathbfcal{Z}} \phi(\mathbfcal{Z}) + \langle \mathbf{\Omega}_t, \mathbfcal{X}_t - \mathbfcal{Z} \rangle + \frac{\lambda_2}{2} \| \mathbfcal{X}_t - \mathbfcal{Z} \|_2^2 \\
&= \argmin_{\mathbfcal{Z}} \phi(\mathbfcal{Z}) - \langle \mathbf{\Omega}_t, \mathbfcal{Z} \rangle + \frac{\lambda_2}{2} \text{Tr}\left( (\mathbfcal{Z} - \mathbfcal{X}_t)(\mathbfcal{Z} - \mathbfcal{X}_t)^T \right) \\
&= \argmin_{\mathbfcal{Z}} \phi(\mathbfcal{Z}) - \langle \mathbf{\Omega}_t, \mathbfcal{Z} \rangle + \frac{\lambda_2}{2} \left( \|\mathbfcal{Z}\|_F^2 - 2\langle \mathbfcal{Z}, \mathbfcal{X}_t \rangle + \|\mathbfcal{X}_t\|_F^2 \right) \\
&= \argmin_{\mathbfcal{Z}} \phi(\mathbfcal{Z}) + \left\langle -\mathbf{\Omega}_t - \lambda_2 \mathbfcal{X}_t, \mathbfcal{Z} \right\rangle + \frac{\lambda_2}{2} \|\mathbfcal{Z}\|_F^2 \\
&= \argmin_{\mathbfcal{Z}} \phi(\mathbfcal{Z}) + \frac{\lambda_2}{2} \left\| \mathbfcal{Z} - \left( \mathbfcal{X}_t + \frac{\mathbf{\Omega}_t}{\lambda_2} \right) \right\|_F^2
\end{aligned}
\end{equation*}

\begin{equation}
   = \mathbfcal{D}_{\phi}\left(\mathbfcal{X}_{t} + \frac{\mathbf{\Omega}_{t}}{\lambda_{2}}\right)
\end{equation} 

The operator $\mathbfcal{D}_{\phi}$ is defined in terms of regularization functions $\phi(\cdot)$. Following the approach outlined in the paper, we extend $\phi(\cdot)$ to capture a wider spectrum of visual features and implement the data operator $\mathbfcal{D}$ using a ResUNet architecture~\cite{diakogiannis2020resunet}. This formulation allows the operator to learn directly from data, where the network parameters $\mathbf{\omega}_{k}$ are updated based on the inputs $\mathbfcal{X}_{t} + \tfrac{\mathbf{\Omega}_{t}}{\lambda_{2}}$, producing the output $\mathbfcal{Z}_{t+1}$.

An analogous expansion is applied when solving for $\mathbfcal{P}_{t+1}$, which leads to the solutions presented earlier in Equation~\ref{eq:pt1}.

\textbf{Update} \(\mathbfcal{X}_{t+1}\):
\nolinebreak
\begin{equation}
\begin{aligned} 
\mathbfcal{X}_{t+1}=\argmin_{\mathbfcal{X}}\langle\mathbf{\Gamma}_{t}, \mathbfcal{H}\otimes\mathbfcal{X}_{t} - \mathbfcal{U}_{t+1}\rangle \\
+ \frac{\lambda_{1}}{2}\|\mathbfcal{H}\otimes\mathbfcal{X}_{t} - \mathbfcal{U}_{t+1}\|_{2}^{2} + \langle\mathbf{\Omega_{t}}, \mathbfcal{X}_{t} - \mathbfcal{Z}_{t+1}\rangle \\+ \frac{\lambda_{2}}{2}\|\mathbfcal{X}_{t} - \mathbfcal{Z}_{t+1}\|_{2}^{2}
\end{aligned}
\end{equation}

\begin{equation}
        \Leftrightarrow \mathbf{\Gamma}_{t}\mathbfcal{H}^{T} + \lambda_{1}(\mathbfcal{H}\mathbfcal{H}^{T}\mathbfcal{X}_{t} - \mathbfcal{U}_{t+1}\mathbfcal{H}^{T}) + \mathbf{\Omega_{t}} + \lambda_{2}(\mathbfcal{X}_{t} - \mathbfcal{Z}_{t+1}) = 0
       \label{eq:xt}
\end{equation} 

To solve Equation (\ref{eq:xt}), the Fast Fourier transform (FFT) can
be utilized, similar to \cite{zhang2020deep}, and \(\mathbfcal{X}_{t+1}\) can be computed as follow:

\begin{equation}
       \mathbfcal{X}_{t+1} = \mathbfcal{F}^{-1}\left\{\frac{\mathbfcal{F}(\mathbfcal{H}^{T}(-\mathbf{\Gamma}_{t} + \lambda_{1}\mathbfcal{U}_{t+1}) - \mathbf{\Omega}_{t} + \lambda_{2}\mathbfcal{Z}_{t+1})}{\lambda_{1}\mathbfcal{F}(\mathbfcal{H})^{2} + \lambda_{2}}\right\}
\end{equation} 
\end{document}